\documentclass[aps,pra,twocolumn,superscriptaddress,longbibliography,
amsmath,amssymb,groupaddress,
]{revtex4-2}
\usepackage[colorlinks,allcolors=blue]{hyperref}

\usepackage{float}
\usepackage{physics,xcolor,palatino}
\usepackage{graphicx}
\usepackage{dcolumn}
\usepackage{bm}
\usepackage{tikz}
\usepackage{changepage}
\usetikzlibrary{shapes.geometric, arrows,positioning}
\tikzstyle{startstop} = [rectangle, text width=4.5cm, minimum   height=1cm,text justified, draw=black, fill=red!30]
\tikzstyle{io} = [trapezium, trapezium left angle=70, trapezium right 
angle=110, text width=3cm, minimum height=1cm, text centered, draw=black, fill=blue!30]
\tikzstyle{process} = [rectangle, text width=4.5cm, minimum height=1cm,    text justified, draw=black, fill=orange!30]
\tikzstyle{processsmall} = [rectangle, text width=3cm, minimum height=1cm,    text justified, draw=black, fill=blue!30]
\tikzstyle{decision} = [diamond, text width=3cm, minimum height=1cm, text centered, draw=black, fill=green!30]
\tikzstyle{arrow} = [thick,->,>=stealth]

\usepackage[normalem]{ulem}

\usepackage[T1]{fontenc}  
\usepackage{ulem}

\makeatletter
\newcommand{\thickhline}{%
    \noalign {\ifnum 0=`}\fi \hrule height 1pt
    \futurelet \reserved@a \@xhline
}
\newcolumntype{"}{@{\hskip\tabcolsep\vrule width 1pt\hskip\tabcolsep}}
\makeatother

\begin{document}

 \title{Inferring Quantum Network Topologies using Genetic Optimisation of Indirect Measurements}

\author{Conall J. Campbell}
\affiliation{Centre for Quantum Materials and Technologies, School of Mathematics and Physics, Queen’s University Belfast, BT7 1NN Belfast, UK}

\author{Matthew Mackinnon}
\affiliation{Centre for Quantum Materials and Technologies, School of Mathematics and Physics, Queen’s University Belfast, BT7 1NN Belfast, UK}

\author{Mauro Paternostro}
\affiliation{Universit\`a degli Studi di Palermo, Dipartimento di Fisica e Chimica - Emilio Segr\`e, via Archirafi 36, I-90123 Palermo, Italy}
\affiliation{Centre for Quantum Materials and Technologies, School of Mathematics and Physics, Queen’s University Belfast, BT7 1NN Belfast, UK}

\author{Diana A. Chisholm}
\affiliation{Universit\`a degli Studi di Palermo, Dipartimento di Fisica e Chimica - Emilio Segr\`e, via Archirafi 36, I-90123 Palermo, Italy}

\date{\today}

\begin{abstract}

The characterisation of quantum networks is fundamental to understanding how energy and information propagates through complex systems, with applications in control, communication, error mitigation and energy transfer.
In this work, we explore the use of external probes to infer the network topology in the context of continuous-time quantum walks, where a single excitation traverses the network with a pattern strongly influenced by its topology. The probes act as decay channels for the excitation, and can be interpreted as performing an indirect measurement on the network dynamics. By making use of a Genetic Optimisation algorithm, we demonstrate that the data collected by the probes can be used to successfully reconstruct the topology of any quantum network with high success rates, where performance is limited only by computational resources for large network sizes. Moreover, we show that increasing the number of probes significantly simplifies the reconstruction task, revealing a tradeoff between the number of probes and the required computational power.

\end{abstract} 

\maketitle

\section{Introduction}
\label{sec_introduction}

Quantum networks provide a powerful platform for distributing, processing, and storing quantum information across multiple interconnected systems, playing a significant role in the advancement of emerging quantum technologies~\cite{Paper11,Paper12,Paper14,Paper17,Paper35,Paper38,Paper39,Paper40,Paper41,Paper42,Paper43, ballarin2025driving, Paper48}. Reconstructing the topology of unknown networks is therefore a fundamental task for quantum technological applications. Continuous-time quantum walks (CTQWs)~\cite{10.1116/5.0190168,Chisholm_2021} are an effective tool for modelling the topology-driven dynamics in complex quantum networks. They have been successfully used to investigate the dynamics of transport phenomena in numerous systems~\cite{CTQW_transport_dynamics,CTQW_honeycomb}. A key example is the demonstration of efficient energy transfer in the Fenna-Matthews-Olson complex~\cite{Mohseni_2008}, one of the most extensively studied light-harvesting protein-pigment complexes~\cite{Paper28,Cogdell_Gall_Köhler_2006,Paper44}.
The complete characterisation of a network's topology can be achieved using methods such as quantum process tomography~\cite{Chuang_1997}, which requires however direct measurement of the state of the entire network, with the number of required measurements scaling exponentially with the system's dimension. As a result, such an approach quickly becomes infeasible for even moderate network sizes~\cite{Nielsen_Chuang_2010}.

This work tackles a fundamental challenge of inferring the topology of an unknown quantum network~\cite{Paper50} by leveraging indirect system measurements using a probe. Approaches based on direct measurements, such as ~\cite{10.1116/5.0190168} provide alternative methods for topology reconstruction for up to 10 node networks with noisy direct measurements. The novelty of our approach is that it provides a non-invasive framework for interacting with complex quantum systems without the need for direct access to the network itself. Such probing procedure consists of injecting an excitation into a network node and attaching an external sink to another one, which we model as an irreversible decay channel. The excitation crosses the network to decay into the sink. By measuring the sink, we collect data driven by the underlying network topology. We thus seek to exploit this relationship to infer the topology of the network.

We frame the task as an optimization problem with the goal of constructing a network with measurement data as similar to the target network as possible. As the parameters of this problem are the connections between network nodes, the parameter space is discretised and so lends itself to an approach such as a genetic algorithm (GA)~\cite{10.1116/5.0190168, 10.5555/534133}, a machine learning method inspired by natural evolution. With sufficient computational resources, this algorithm can reconstruct any network topology, and in our case it has the ability to reconstruct the topology of networks up to 10 nodes with high fidelity. Furthermore, by altering the probing procedure to consider injecting the excitation into different nodes, we can simplify the reconstruction procedure and enable a significant reduction in the required computational power. This facilitates reconstruction of topologies for larger networks when constrained by the available computational resources.

The remainder of this manuscript is structured as follows. In Sec.~\ref{sec_the_model} we introduce the physical model under study. In Sec.~\ref{sec_geneticalgorithm} we discuss the GA used to infer the network's topology. Finally, in Sec.~\ref{sec_results} we perform an analysis to determine the optimal algorithm configuration, and report on its performance. Sec.~\ref{sec_conclusions} is devoted to a summary of the main results that we have achieved and provides a further look into interesting directions for future investigation.

\section{The Model}
\label{sec_the_model}

\begin{figure}
    \centering
    \includegraphics[width=1\linewidth]{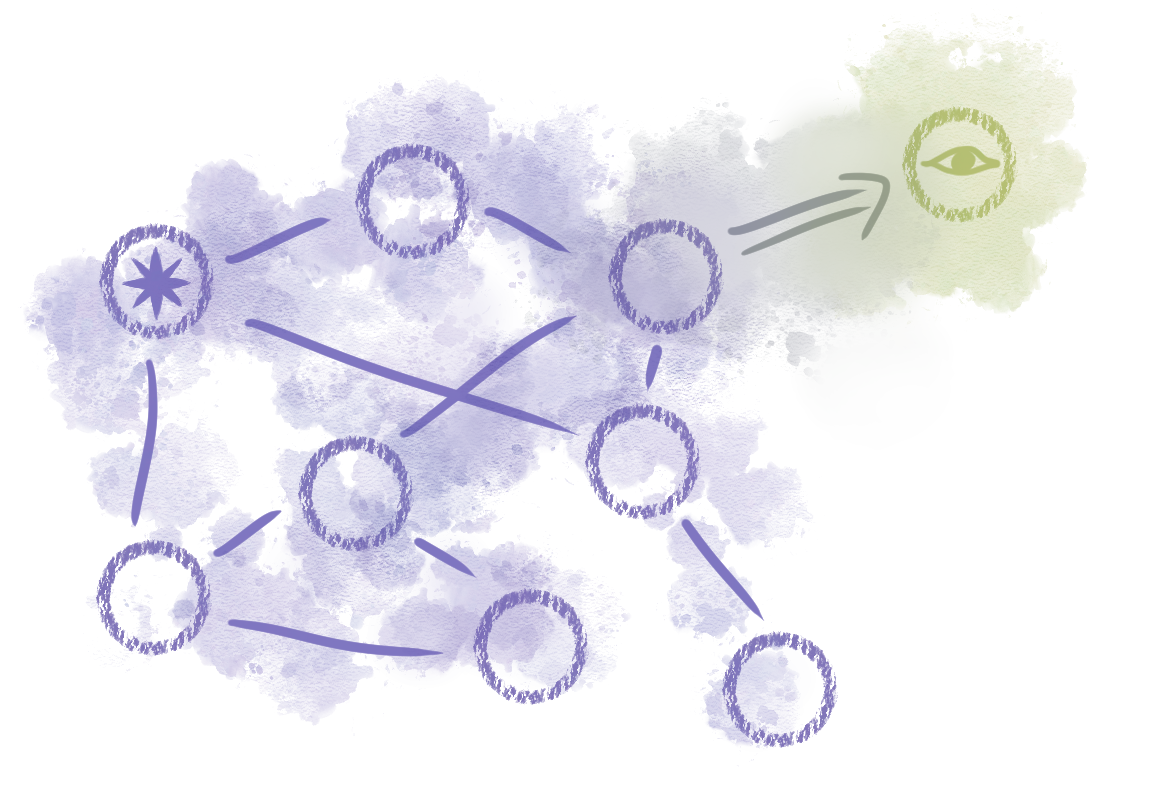}
    \caption{Pictorial representation of a quantum network of interconnected nodes, with the excitation (purple star) being initially localised in one node, and an externally attached sink which is used to perform indirect measurements (green eye)}
    \label{fig:CTQWdiagram}
\end{figure}

The system is modelled as a collection of $n$ qubits (nodes) that form a network which is connected by edges that allow for energy transfer between nodes. The system evolves under the excitation-preserving Hamiltonian 
\begin{equation}
    \label{eq_hamiltonian1}
    H = \sum_{i}^{n}\omega\text{ }\sigma_{i}^{+}\sigma_{i}^{-} + \sum_{i,j}^n\text{ }g_{ij}\text{ }\Big(\sigma_{i}^{+}\sigma_{j}^{-} + \sigma_{j}^{+}\sigma_{i}^{-}\Big),
\end{equation}
which describes the transfer of energy through a quantum network of coupled qubits, with $\sigma_{i}^{\pm}$ the ladder operators of qubit $i$. The first term in Eq.~\eqref{eq_hamiltonian1} represents the on-site energy of node $i$, where we assume each qubit has the same on-site energy $\omega$. The second term represents how the excitation hops from one node to another with hopping strength $g_{ij}$, which we assume for simplicity to only take values 0 and 1. We consider the case where only one excitation is traversing the network, thus restricting the dynamics to the single excitation subspace, described by the single excitation Hamiltonian $H^{(1)}$ with elements
\begin{equation}
    \label{eq_hamiltonian2}
        H^{(1)}_{ij} = \langle i|H|j\rangle = g_{ij}(1-\delta_{ij})+\omega\delta_{ij},
\end{equation}
with $\delta_{ij}$ the Kronecker symbol. We thus have that 
\begin{equation}
    H^{(1)}_{ij} =
                    \begin{cases} 
                        \omega & \text{if } i = j\\
                        g_{ij} & \text{if } i \neq j
                    \end{cases},
\end{equation}
where $\{|i\rangle: 0\leq i < n\}$ is the basis for the single excitation subspace with $|i\rangle$ representing the physical configuration where the excitation is localised at node $i$. Notably, $H^{(1)}$ significantly reduces the complexity of computation, as the dimension of the Hilbert space now scales linearly with the number of nodes, rather than exponentially. As the  on-site operators commute with the total Hamiltonian, they do not contribute to the dynamical evolution of the system. We thus define the adjacency matrix $A$ as
\begin{equation}
    \label{adjacency matrix}
    A_{ij} = (1-\delta_{ij})H_{ij}^{(1)},
\end{equation}
which is a symmetric $n\times n$ matrix that encodes the topology of the network. Thus, to reconstruct the network topology, we aim to estimate $A$ by monitoring the system's dynamics. However, rather than directly measuring the state of the network to determine the topology, we probe its excitation dynamics by externally attaching a sink to the network nodes~\cite{Chisholm_2021}. A sink attached to a node $s$ acts as a decay channel, which can be modelled by means of an effective non-hermitian Hamiltonian $H_{sink}$ defined as
\begin{equation}
    \label{eq_hamiltonian3}
    H_{sink} = A-i\gamma|s\rangle\langle s|,
\end{equation}
with $\gamma$ the decay rate. This mechanism mirrors the function of the reaction centre in the FMO complex, where absorbed light is irreversibly converted into chemical energy. In the model, once the excitation reaches node $s$, it exponentially decays into the sink at a rate controlled by $\gamma$, and cannot re-enter the network once absorbed. A schematic of the network and external sink probe is shown in Fig. \ref{fig:CTQWdiagram}

The initial state of the network is chosen as a single excitation injected at the $e^{\text{th}}$ node, and we denote $\rho(t)$ as the state of the system at time $t$, evolving under the Hamiltonian in Eq.~\eqref{eq_hamiltonian3}.  The non-Hermiticity of $H_{sink}$ results in dynamics that are no longer excitation-preserving, so that $\rho(t)$ is no longer of trace one. The population loss of the network is accounted for by the increase in the sink population, so that the sum of the two is always one. We have therefore that the population of the sink at time $t$ is given by

\begin{equation}
    \label{sink population}
    P_s(t) = 1- \text{Tr}[\rho(t)].
\end{equation}
The dynamics of the sink population are directly dependent on the complete structure of $A$, however the initial evolution will be dominated by the shortest paths between the initial excitation node and the sink node.

A Taylor series expansion of the propagator truncated at second order yields an expression for the state of the system at time $t$ as 
\begin{equation}
    \label{Taylor_series_eq}
    |\psi(t)\rangle = e^{-iH_{sink}^{(1)}t}|e\rangle\approx \sum_{n=0}^2\frac{(it)^n}{n!}(A-i\gamma|s\rangle\langle s|)^n|e\rangle.
\end{equation}
Focusing on the case where the sink and excitation are located at different nodes in the network, $\langle s|e\rangle=0$, Eq.~\eqref{Taylor_series_eq} simplifies to the form
\begin{equation}
    |\psi(t)\rangle \approx |e\rangle + itA|e\rangle -\frac{t^2}{2}A^2|e\rangle +\frac{i\gamma t^2}{2} A_{es}|s\rangle.
    \label{short time evolution}
\end{equation}
Combining Eq.~\eqref{sink population} and Eq.~\eqref{short time evolution}, we have
the approximate expression for the short-time population of the sink 
\begin{equation}
    \label{short time sink population}
    P_s(t)\approx \gamma A_{es}t^3-\left[\gamma A_{es} + \sum_i\langle i|A^2|e\rangle^2\right]\frac{t^4}{4}.
\end{equation}

For $t\ll 1$, the sink population depends primarily on $A_{es}$ and so the short-time sink population can be used to directly infer whether nodes $e$ and $s$ are connected. The elements of the adjacency matrix describing connections between other nodes are correlated with the sink population at longer times and exhibit complex non-linear relationships, requiring more advanced techniques for their estimation.

\section{The Genetic Algorithm}
\label{sec_geneticalgorithm}

GAs allow for the optimisation of problems with complex high-dimensional search spaces. A GA utilises iterative update methods inspired by the evolutionary principles of selection, crossover, and mutation on candidate solutions. In particular, GAs excel at optimisation problems with discretised search spaces, where traditional optimisation techniques such as stochastic gradient descent are not applicable~\cite{10.5555/534133}. 

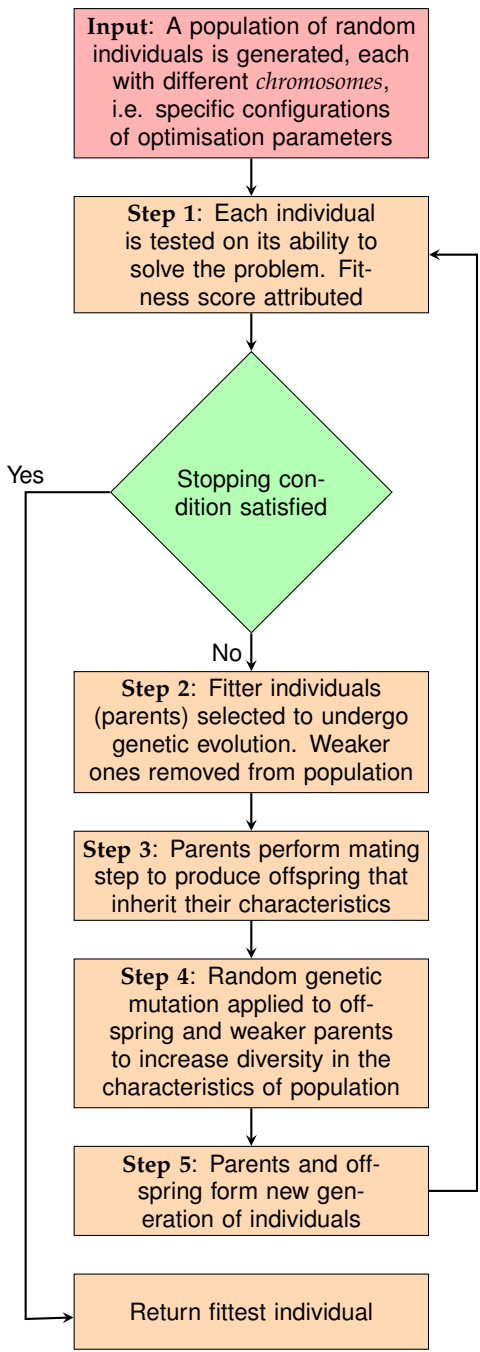
\begin{figure}
\begin{adjustwidth}{-2.5cm}{-2.5cm}
  \begin{center}
    \begin{tikzpicture}[node distance=0.5cm,font=\sf]
      \node (in1) [startstop, align=center] {\small{{\bf Input}: A population of random individuals is generated, each with different {\it chromosomes}, i.e. specific configurations of optimisation parameters}};
      \node (pro1) [process, below=of in1, align=center] {\small{{\bf Step 1}: Each individual is tested on its ability to solve the problem. Fitness score attributed}};
      \node (decision) [decision, below=of pro1, align=center, inner sep=1pt] {Stopping condition satisfied};
      \node (pro2) [process, below=of decision, align=center] {\small{{\bf Step 2}: Fitter individuals (parents) selected to undergo genetic evolution. Weaker ones removed from population}};
      \node (pro3) [process, below=of pro2, align=center] {\small{{\bf Step 3}: Parents perform mating step to produce offspring that inherit their characteristics}};
       \node (pro4) [process, below=of pro3, align=center] {\small{{\bf Step 4}: Random genetic mutation applied to offspring and weaker parents to increase diversity in the characteristics of  population}};
       \node (pro5) [process, below=of pro4, align=center] {\small{{\bf Step 5}: Parents and offspring form  new generation of individuals}};
       \node (finish) [process, below=of pro5, align=center] {Return fittest individual};
      
      \draw [arrow] (in1) -- (pro1);
      \draw [arrow] (pro1) -- (decision);
      \draw [arrow] (decision) -- node[anchor=east] {No} (pro2);
      \draw [arrow] (pro2) -- (pro3);
      \draw [arrow] (pro3) -- (pro4);
      \draw [arrow] (pro4) -- (pro5);
      \draw [arrow] (pro5) -| ++(3,1) |-  (pro1);
      \draw [arrow] (decision)  -- ++(-3,0) node[anchor=south] {Yes}     
              -- ++(0,-9)     
              -- ++(0,-2)      
              --  (finish) ;
    \end{tikzpicture}
  \end{center}
\end{adjustwidth}
\caption{General workflow of a GA.}
\label{genericworkflow}
\end{figure}

A scheme of principle for the structure of a GA is as in Fig.~\ref{genericworkflow}.

In our work, the target and population networks are chosen as $n$-node Erdös-Rényi graphs~\cite{ERgraphs}, where each edge connection is included with random probability $p\in[0,1]$, resulting in networks of all different edge densities. Furthermore, we consider only connected graphs. The sink population data for every network is collected according to the following \mbox{procedure}: 
\begin{enumerate}
    \item An excitation is injected at node $e$ of the network while the sink is placed at node $s$;
    \item The system is allowed to evolve for some time $t$;
    \item $P_s(t_i)$ is measured at times $\{t_i\}$ for $i$ time-steps;
    \item The system is reinitialised, and the probing procedure repeated for each possible sink location.
\end{enumerate}

As discussed in Sec.~\ref{sec_the_model}, it is possible to infer, using the short-time sink population, whether there is a direct connection between nodes $e$ and $s$, regardless of the location of the latter. 
We thus set known adjacency matrix elements to match that of the target for all members of the initial population and offspring generated by the GA. This reduces the number of optimisation parameters from ${n(n-1)}/{2}$ to ${(n-1)(n-2)}/{2}$.

\begin{figure*}
    \centering
        \includegraphics[width=1
        \textwidth]{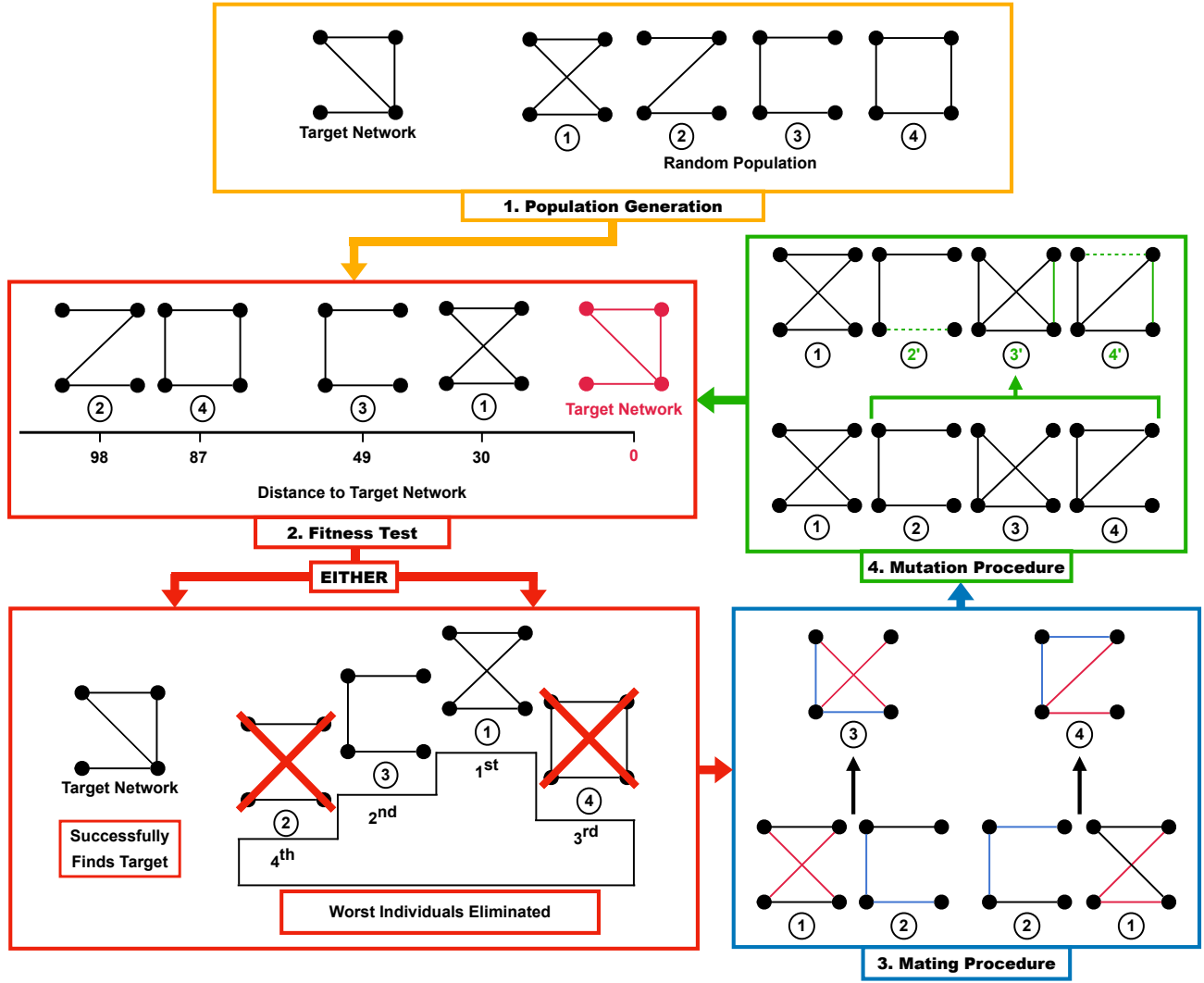}
    \caption{Flow chart of GA. \textbf{1.} Initial population and target individual generated. \textbf{2.} Evaluation of fitness function and ordering of individuals from best to worst. {\it Either} the algorithm finds the target network and stops, or weaker individuals are eliminated and stronger individuals selected for genetic evolution. \textbf{3.} Mating procedure, two selected individuals contribute chromosomes to form offspring. \textbf{4.} Slight genetic mutations are induced in weaker parent and all offspring, adding or removing chromosomes from the population, potentially resulting in new characteristics. The {new generation} has their fitness evaluated and the algorithm loops until a target solution is found or the algorithm reaches the pre-set iteration limit.}
    \label{fig:GA_flowchart}
\end{figure*}

Each network $A^{j}$ receives a fitness score $\mathcal{F}_j$ 
, based on the difference between its sink population and the sink population of the target network $A^{\text{targ}}$ at times $\{t_i\}$, that is 
\begin{equation}
    \mathcal{F}_{j} = \sum_{i}\sum_s\left|P_s^{(A^{\text{targ}})}(t_i) - P_s^{(A^{j})}(t_i)\right|.
    \label{fitness function}
\end{equation}
We make the heuristic, yet reasonable assumption that similarity in sink population reflects similarity in the adjacency matrices of two networks. Hence, the parents are selected as the fittest 20\% of the population, which corresponds to those individuals with the smallest fitness score $\mathcal{F}_j$. Since these parents survive into the new generation, this is known as an {\it elitist} strategy. This aims to accelerate the convergence of the algorithm by maintaining the highest quality solutions at each iteration~\cite{elitism}. A flow chart of the algorithm is presented in Fig.~\ref{fig:GA_flowchart}, specializing the generic methodology illustrated in Fig.~\ref{genericworkflow} to the network-based case addressed here. 

The parents then undergo the mating procedure, in order to restore the original population size. Each parent mates with four randomly selected parents. The mating of parents $a$ and $b$ proceeds as follows
\begin{enumerate}

    \item A mating probability is evaluated as $p_{mat} = \frac{{\mathcal{F}_B}}{{\mathcal{F}_A}+{\mathcal{F}_B}}\in[0,1]$ using their respective fitness scores. This ensures that the fitter individual will pass on more of their chromosomes to the offspring.
    \item A new offspring adjacency matrix $A$ is constructed by taking element $A_{ij}=A_{ji}$ from parent $a$ with probability $p_{\text{mat}}^{a}$ and parent $b$ otherwise.
\end{enumerate}
Following the mating procedure, the weaker half of the parents and all offspring will undergo mutation. Each element of their adjacency matrices is flipped with a probability of mutation sampled from an exponential probability distribution. The distribution average is $p_{\text{mut}}=\frac{1}{\alpha}$ where $\alpha$ is the number of optimisation parameters~\cite{HE2002245}. Here, $\alpha$ is the total number of connections to be estimated, which ensures that on average one connection is changed.

Due to the non-linear relationship between sink data and the adjacency matrix, the search space contains many local minima in which the algorithm can get stuck. We thus make two modifications to the algorithm to assist it when it encounters said minima. The first modification monitors whether the fittest 10\% of the population (unmutated parents) stays the same throughout multiple iterations. If this population does not change after 20 iterations, then $p_{\text{mut}}$ is increased to 0.1 for a single iteration, causing on average 10\% of the connections to be change (extreme mutation), facilitating a large change of trial solutions. The second modification considers the case where a total of 5 extreme mutations are performed. The weakest 20\% of the population are replaced with a random set of individuals, thus injecting potentially unconsidered networks into the population, allowing us to explore a different area of the search space.

A scenario that may occur during the execution of the algorithm is that a network in the population is an isomorphism of the target network, meaning it is identical to the target network up to a simple relabelling of its nodes: when comparing two isomorphic networks, the fitness $\mathcal{F}$ will be non-zero, even though they have an identical topology. It is possible to identify a pair of isomorphic networks based on their sink populations by searching for the isomorphism between the two sets of sink population data. However, the computational complexity of such a task scales with the network's size as $O(n\log n)$~\cite{sorting_algs}, making it intractable for large networks. We can however quickly check a necessary condition for two networks to be isomorphic by computing the difference in the sum over each network's sink data
\begin{equation}
    \Delta_j=\sum_{t_i,s}\left|P_s^{(A^{\text{targ}})}(t_i)\right| -  \sum_{t_i,s}\left|P_s^{(A^{j})}(t_i)\right|.
\end{equation}
Here, $\Delta_j$ is non-zero if and only if the two networks are not isomorphic. To identify isomorphic networks, a two step procedure is implemented. First, networks for which $\Delta_j=0$ are identified. For these, a sorting algorithm is used to check for an isomorphism. Should one be found, the algorithm is halted.

\section{Results}
\label{sec_results}

In Sec.~\ref{fixed_excitation_node_perfromance_sec}, 
we analyse the effect that the algorithm's controllable parameters have on its performance, in order to find the optimal setup, before determining the convergence performance of the GA. Finally, in Sec.~\ref{variable_excitation_convergence_performance_sec} we relax the constraint on the excitation being injected into the same node of the network, and explore the potential to improve performance for larger networks by increasing the amount of data collected.

\subsection{Optimal Model Parameters and Performance}
\label{fixed_excitation_node_perfromance_sec}

Quantifying the difficulty of a given search problem remains an open issue in the field of GAs. A promising metric for estimating solvability is the correlation of the trial solution fitness $\mathcal{F}$ with a measure of the true distance $\mathcal{D}$ from the global optimum, referred to as Fitness Distance Correlation (FDC)~\cite{fitness_evaluation}. This is defined as the Pearson's correlation coefficient $P_C$ between $\mathcal{F}$ and $\mathcal{D}$ for a set of trial solutions to the problem
\begin{equation}
    \text{FDC} = P_C(\mathcal{F},\mathcal{D}) = \frac{\text{cov}(\mathcal{F},\mathcal{D})}{\sigma_\mathcal{F}\,\sigma_\mathcal{D}}
    \label{pearsons correlation}.
\end{equation}

This is a measure of linear correlation between $\mathcal{F}$ and $\mathcal{D}$. A positive (negative) FDC reflects (anti-)correlation. For the minimization problem considered in this work, a larger FDC is indicative of a more solvable problem for a GA, while a negative FDC implies that, on average, the fitness function guides the algorithm away from the true solution. We define the distance $\mathcal{D}$ of a trial adjacency matrix $A'$ from the target adjacency matrix $A^\text{targ}$ to be 
\begin{equation}
    \mathcal{D}(A^\text{targ}, A') =  \frac{1}{2}\sum_{i,j}^N|A^\text{targ}_{ij} - A'_{ij}|
    \label{adjacency_distance}.
\end{equation}
This metric describes the total number of topological differences between the networks, ranging from 0 to $n(n-1)/2$, and thus serves an effective measure of distance between them. Using this metric to compute the FDC, it is possible to estimate the reconstructability of a given network topology by means of a GA. As a result, to maximize the performance of the algorithm, it is necessary to choose the times at which measurements on the sink are performed so as to maximize the average FDC of the problem over the space of possible target networks. 

\begin{figure}
    \centering
    \includegraphics[width=1\linewidth]{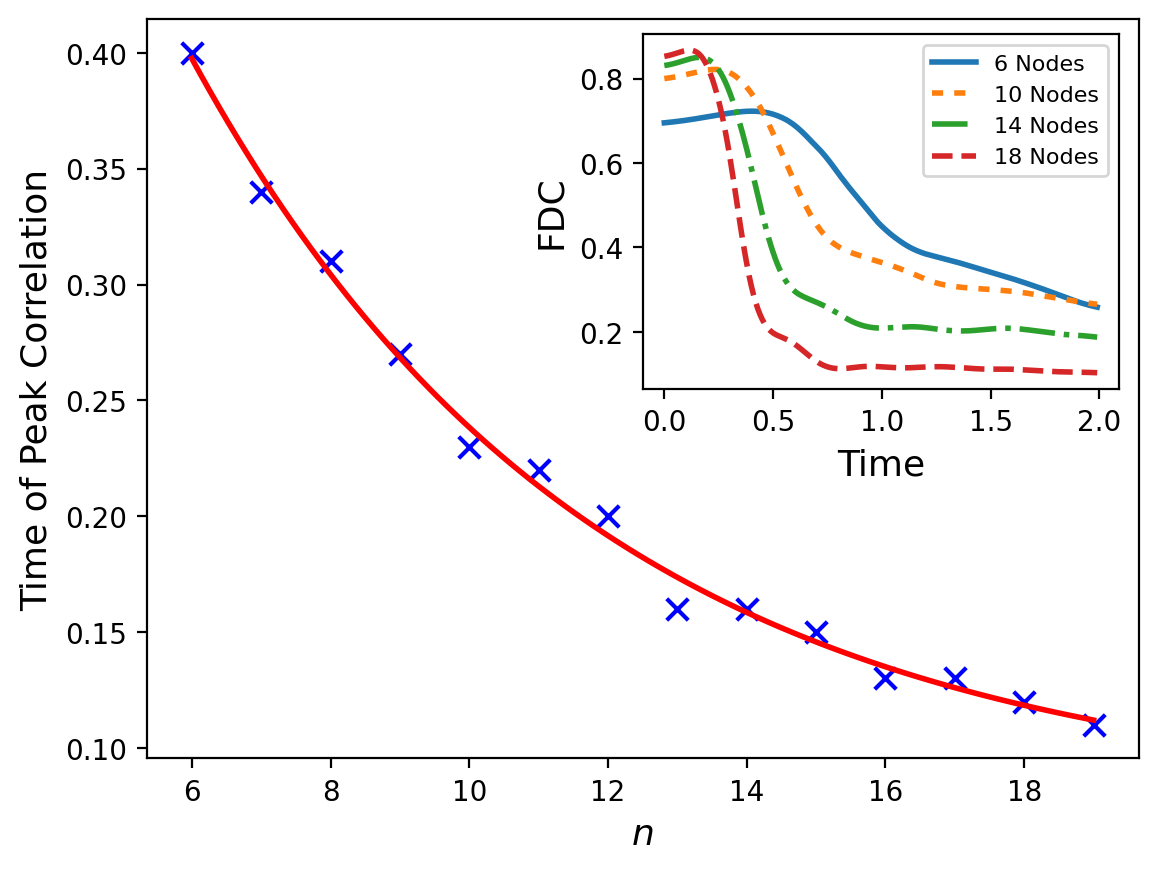}
    \caption{Dynamics of FDC over time for fully random Erdös-Rényi networks. {\bf Main panel}: We show the time at which the maximum value of the FDC is achieved against the number of nodes in a network. The blue crosses represent simulated data, while the red line represents an inverse exponential fit. {\bf Inset}: FDC as a function of time for varying network dimension. 
    }
    \label{fig:time_v_correlation}
\end{figure}

To investigate the average FDC as a function of time, a set of target networks and a set of trial solution networks, each of $n$ nodes, were generated. For each target, the FDC was calculated using the set of trial solution networks, with $\mathcal{F}$ computed using the sink data at each point in time $t$. The average FDC was then computed over the set of target networks. This allows us to find, for a given network size, a time $t$ for which the sink data maximises the FDC. By repeating this procedure for different network sizes, we determine the time at which FDC peaks, as shown in the main panel of Fig.~\ref{fig:time_v_correlation} which shows that FDC decreases as the size of system increases. More specifically, the time of peak correlation follows an exponential decay as shown by the red curve. The inset in Fig.~\ref{fig:time_v_correlation} shows the average FDC as a function of time for various system sizes.
As discussed in Sec.~\ref{sec_the_model}, the short time evolution of the sink population is dominated by whether the initial excitation node and sink node are connected, consequently, the FDC is high at short times before rapidly dropping as the network dynamics become more complex.

\begin{figure}[t!]
    \centering
    \includegraphics[width=1\linewidth]{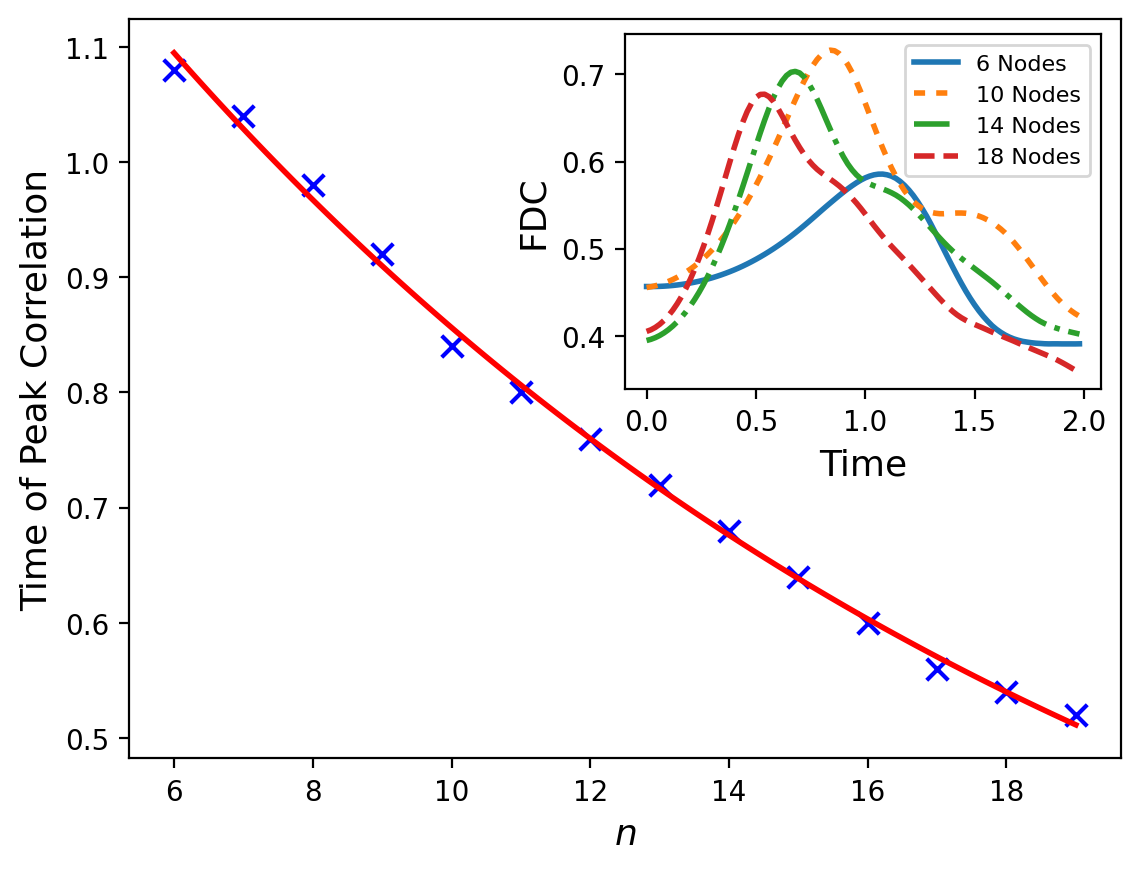}
    \caption{Dynamics of FDC over time for random Erdös-Rényi networks constrained to have matching adjacency matrix elements corresponding to probed pairs of sink and excitation nodes. {\bf Main panel}: Time at which  FDC achieves its maximum as a function of the dimension of the network. Blue crosses represent simulated data, while the red line represents an inverse exponential fitting of the data. {\bf Inset}: FDC as a function of time for varying network dimension.}
    \label{fig:correlation_v_time_fixed}
\end{figure}

As it is possible to infer whether the excitation and sink nodes are connected, the FDC dynamics in Fig.~\ref{fig:time_v_correlation} do not exactly reflect that of the GA setup. When running the algorithm, these connections will always match between the target and the population. As a consequence, it is necessary to consider an alternative setup for the analysis of the FDC. For each target network, a set of trial solution networks were generated with matching adjacency matrix elements where the elements can be inferred from the short-time sink population. Based on this setup, the main panel of Fig.~\ref{fig:correlation_v_time_fixed} shows the time at which FDC achieves its maximum as a function of the number of nodes, with the inset showing the FDC as a function of time for different network sizes.  In contrast to Fig.~\ref{fig:time_v_correlation}, the FDC has both smaller peak values and reaches maximal correlation at a later time than in the previous case. This is due to the non-linear relationship between the sink population and the adjacency matrix elements that do not directly connect the sink and the excitation nodes, which become relevant at longer time scales. Once again, the time of peak correlation follows an inverse exponential relationship with system size. Based on this analysis, the sink population will be measured at times centred on the time of maximum FDC. 

\begin{figure}
    \centering
    \includegraphics[width=\linewidth]{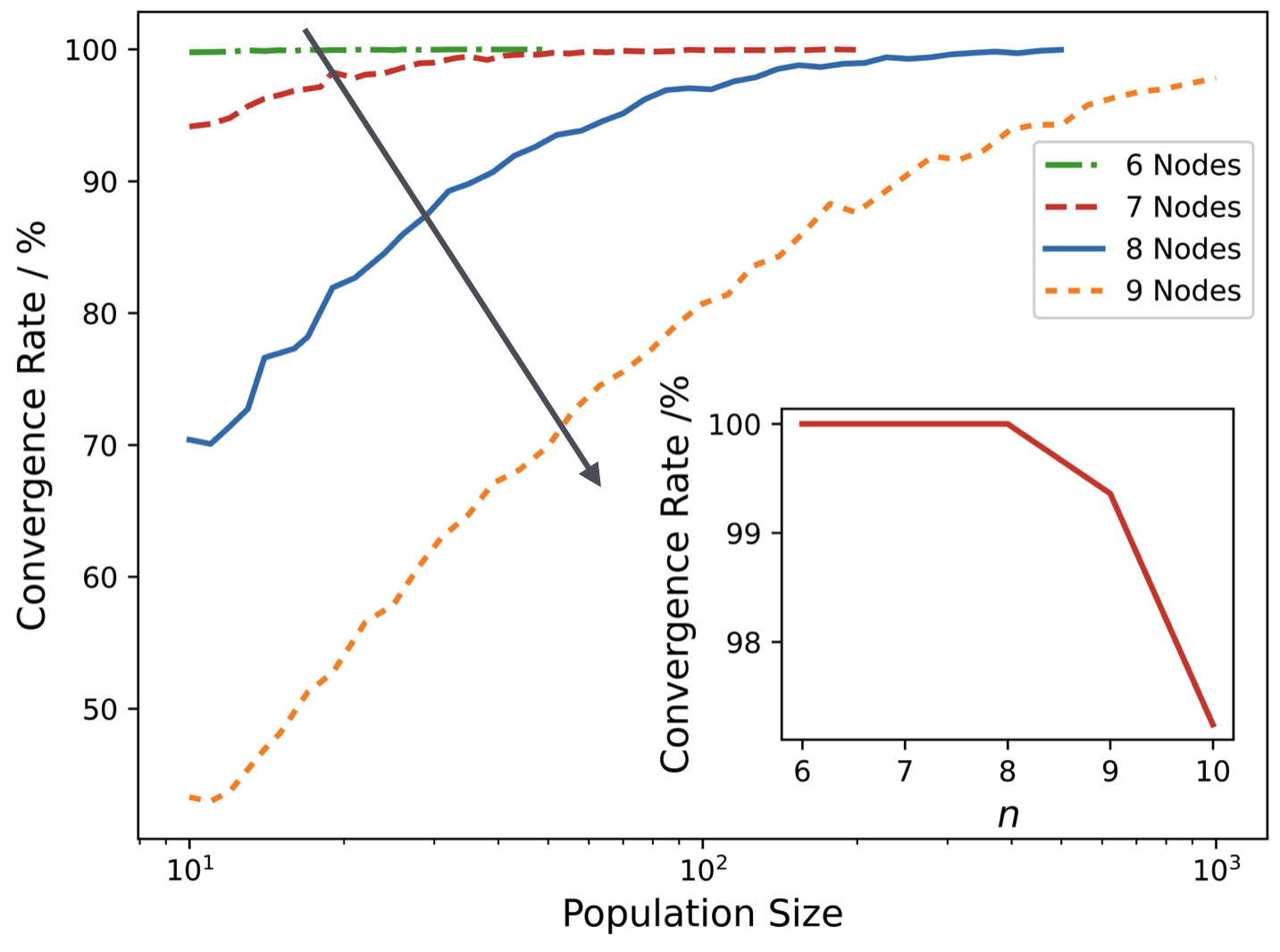}
    \caption{Convergence of the GA for different network sizes. {\bf Main panel}: Convergence rate as a function of the population size for varying network sizes. The black arrow shows the direction of growth of $n$. {\bf Inset}: Number of instances of convergence of the algorithm to the exact target topology for $n=\{6,7,8,9,10\}$ node networks and population sizes $\{100 , 500 , 1000 , 2000 , 10000\}$ respectively.}
    \label{fig:population_size_analysis_plot}
\end{figure}

Beside the solvability of a given problem by a GA, the main driver of the algorithm performance is the size of the population that is used. The population is the resource that the algorithm uses to explore the solution-space of the problem. Therefore, increasing the size of the population is expected to improve the performance of the algorithm, as this provides a better spanning of the solution space. Fig.~\ref{fig:population_size_analysis_plot} shows the performance of the algorithm as a function of the population size for different network dimensions. It can be seen that the size of the population required to achieve optimal performance grows exponentially with the number of network nodes, due to the exponential increase in the size of the search space.

Using the previously discussed parameter configuration, the algorithm performance was assessed by estimating the topology of 5000 target networks for system sizes between 6 and 10 nodes. The rate of successful convergence to the target topology is shown as a function of system size in the inset of Fig.~\ref{fig:population_size_analysis_plot} with population sizes $\{100 , 500 , 1000 , 2000 , 10000\}$ respectively. The model is capable of correctly reconstructing all target networks up to a size of 8 nodes and achieved a performance of over 97\% for $n=9$ and 10. Networks that the algorithm failed to reconstruct were predominantly of a medium connectivity, centered around the area of the search space with a large number of permutations for a given number of connections. As a result, the model is likely to perform better for networks that are either sparsely or densely connected.

The model was ran on a laptop utilising a 10-core CPU with a  3.5GHz clock-speed and 16GB RAM. For networks of 10 nodes, the algorithm had an average runtime per iteration of approximately 2.3s, with an average of 25.5 iterations leading to a total average runtime of roughly 58s. The algorithm utilised roughly 2GB of working memory. The runtime and memory requirements for the algorithm scales close to linearly with the number of network nodes fixing the population size. The limiting factor therefore, is the exponentially increasing population size necessary to reconstruct a target network as the number of nodes is increased. Increasing memory requirements can be handled by batch processing portions of the population sequentially, and so the limiting factor rapidly becomes the runtime of the algorithm for larger system sizes.

\subsection{Reconstruction of Larger Networks Using Multiple Injection Locations}
\label{variable_excitation_convergence_performance_sec}

\begin{figure}
    \centering
    \includegraphics[width=1\linewidth]{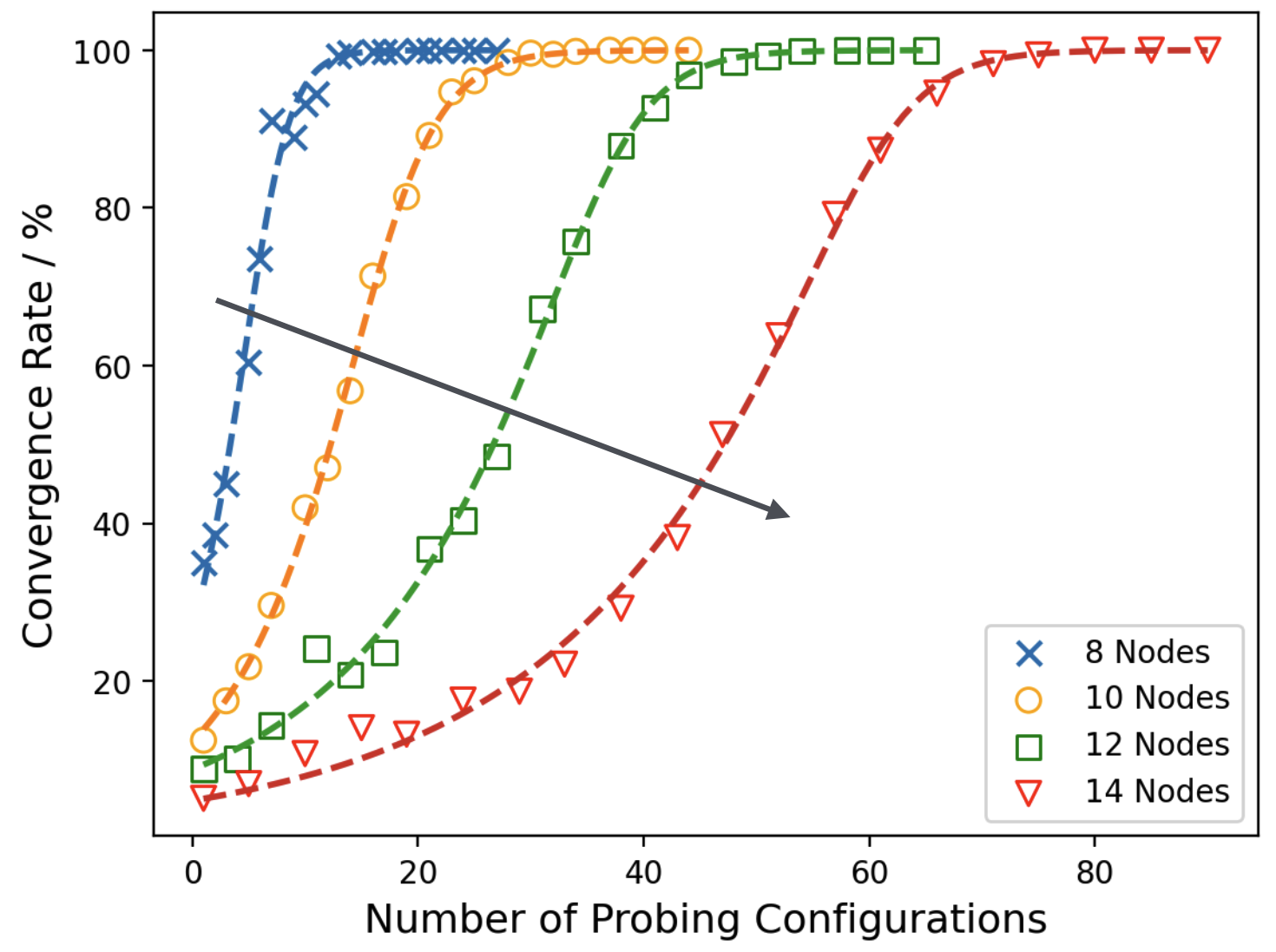}
    \caption{Convergence rate of the GA studied against the number of measurement configurations as the number $n$ of nodes of a network grows. Simulated data is represented with coloured markers, while dashed lines represent a generalised logistic fitting of the data. The black arrow shows the direction of growth of $n$.}
    \label{fig:measurements_vs_performance_fitted}
\end{figure}

As the size of population for which the algorithm can be run quickly becomes a limiting factor for large network size, we devise an alternative approach to scale the algorithm to higher-dimensional networks. Namely, we relax the constraints on the location of the initial excitation injection. By considering the possibility of moving the excitation's initial location, the number of connections to be estimated by the algorithm can be reduced, while the quantity of information available for use in the fitness function is increased. The modified probing procedure is defined as follows:
\begin{enumerate}
    \item An excitation is injected at node $i$ of the network;
    \item The sink is placed at node $j$ of the network;
    \item The system is allowed to evolve for some time $t$;
    \item $P_s(t)$ is measured;
    \item System is reinitialised, and the procedure is repeated for a subset of the set of pairs of all nodes $\{(x,y):x,y<  n\}$.
\end{enumerate}
Through this probing procedure, we can investigate the necessary number of measurements required to achieve optimal performance for a given population size. In Fig.~\ref{fig:measurements_vs_performance_fitted} the performance of the algorithm as a function of the number of probing configurations used is shown for various values of $n$, up to 14 , with a population of 50 networks. The convergence rate was estimated using 1000 sample targets represented by the crosses in the figure. A generalised logistic curve represents an accurate fitting of the data: 
for all network sizes the convergence rate saturates to 100\% given a sufficiently large number of initial measurement configurations. This is to be expected, because if all possible probing configurations are measured, then the structure of the adjacency matrix is explicitly identified by the measurement data. As a result, the problem of reconstructing large networks is reduced to collecting enough measurement data to reduce the size of the problem to that which can be achieved given computational resources.

As the network size increases, the number of extra probing configurations required to reach optimal performance is similar to the increase in the number of connections in the network. This demonstrates that, for a given population size, the model is capable of inferring a roughly constant number of connections. This is due to the complexity of the problem, which is highly dependent on the number of unknown connections to be inferred.
\begin{figure}[t!]
    \centering
    \includegraphics[width=1\linewidth]{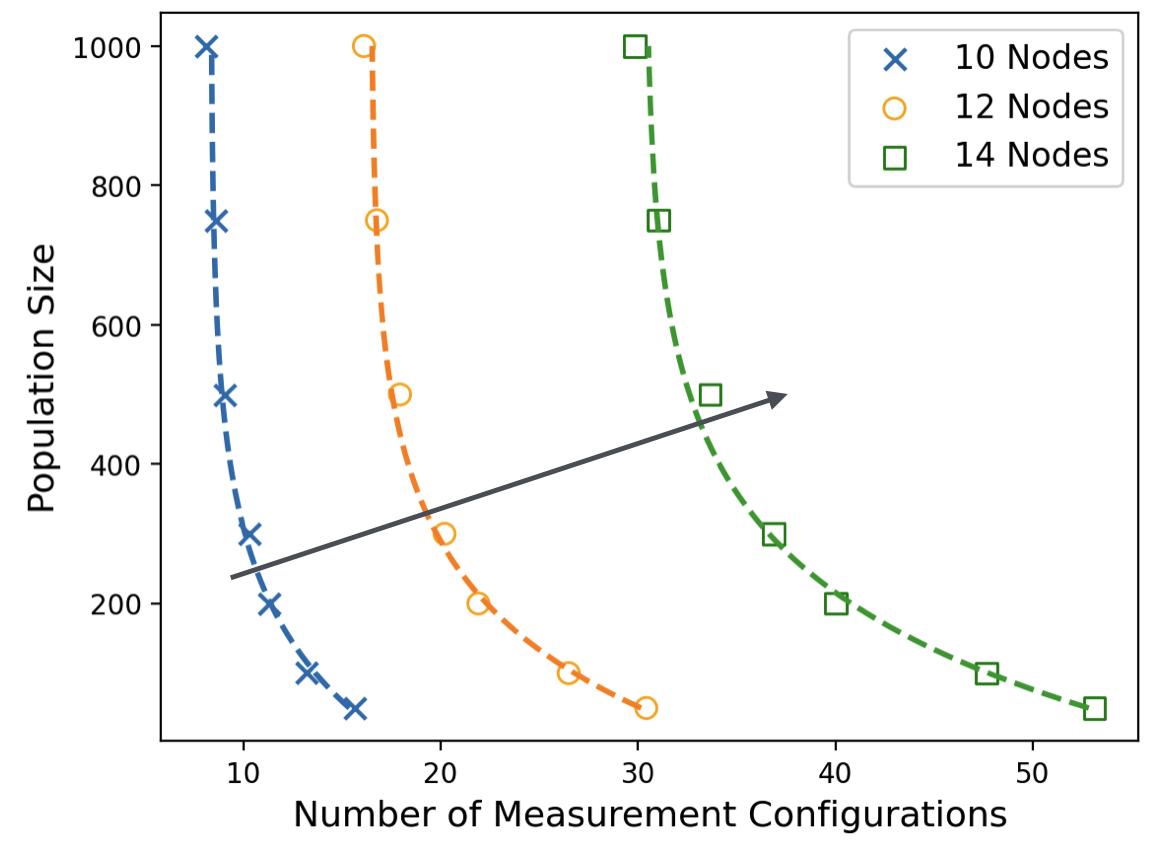}
    \caption{Average population size required for convergence of the algorithm at a given number of measurement configurations for varying system sizes. Simulated data is represented with coloured markers and dashed lines represent a first order rational function fitting of the data. The black arrow shows the direction of growth of $n$.}
    \label{fig:min_pop_size_vs_measure}
\end{figure}

The ability of the algorithm to reconstruct a given network is clearly highly dependent on the size of population and the number of measurement configurations utilized during the measurement procedure. This is displayed in Fig.~\ref{fig:min_pop_size_vs_measure}, which shows the population size needed to guarantee the reconstruction of a given network as a function of the number of measurement configurations being used. By increasing the latter, the necessary population size for network reconstruction decreases, thus providing a method to mitigate the exponential increase in computational resources required as system size increases (Fig. \ref{fig:population_size_analysis_plot}), and making this approach more feasible for simulation on computers with limited resources. This highlights the necessary trade-off between effort invested in collection of sink data, depending on the number of measurement configurations, and the computational resources that are available for the problem. As a result, this method facilitates the reconstruction of any size of network given a number of measurements and a given population size. However, the necessary computational resources and experimental requirements will likely become very demanding for very large network sizes.

\begin{figure}[t!]
    \centering
    \includegraphics[width=1\linewidth]{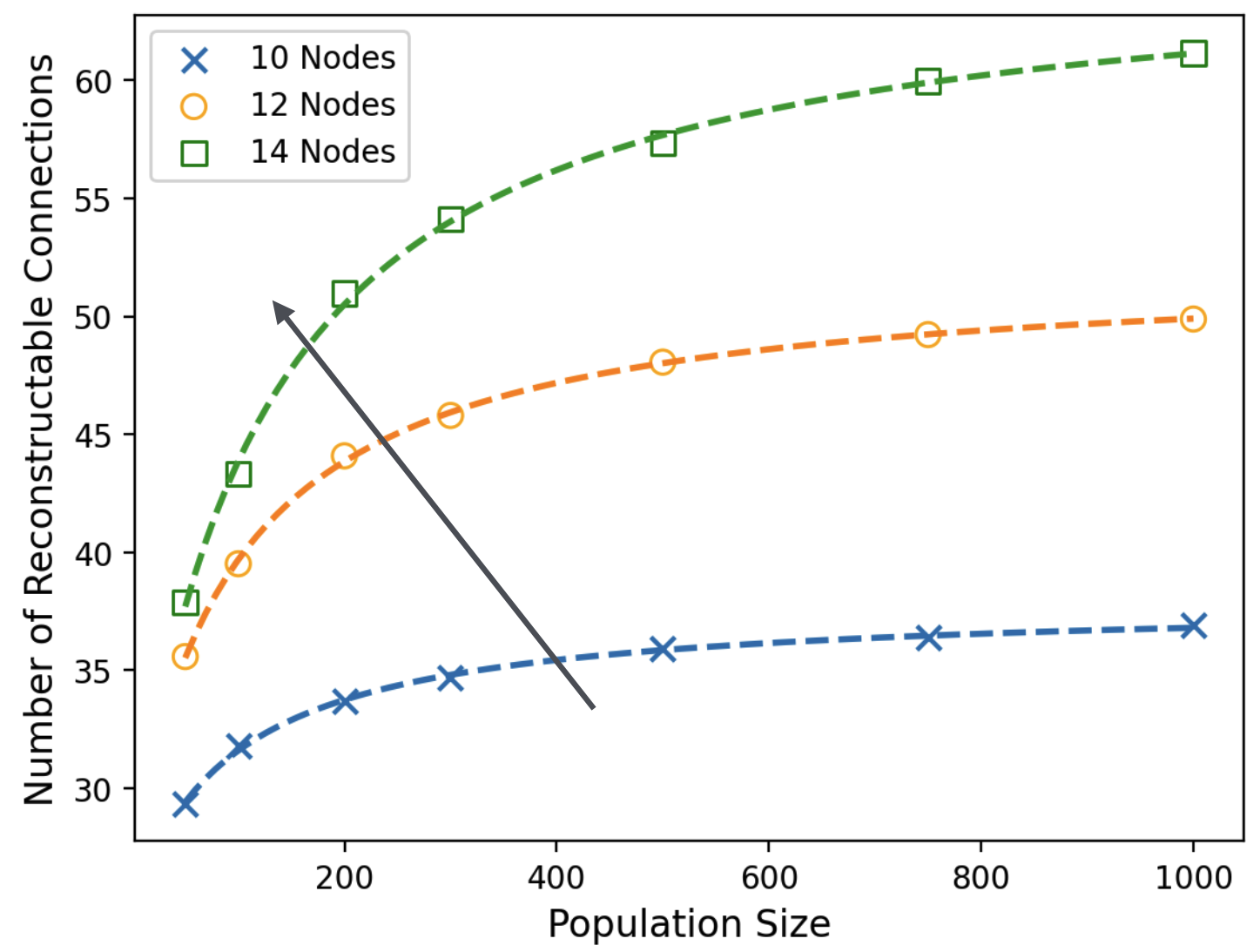}
    \caption{Average number of unknown network connections that can be successfully reconstructed as a function of population size in the algorithm for various sizes of target network. Simulated data represented by coloured markers and dashed lines represent a first order rational function fitting of the data. The black arrow shows the direction of growth of $n$}
    \label{fig:number_of_reconstructable_connections}
\end{figure}

While the complexity of the problem increases as the size of the network grows, an interesting observation is that the algorithm is also capable of inferring an increasing number of unknown connections. Fig.~\ref{fig:number_of_reconstructable_connections}  shows the maximum number of unknown connections for which the model is capable of converging at different sizes of population. This is shown to increase along with the size of the network, possibly due to a greater number of measurement configurations being used. This results in a larger amount of information being collected about the dynamics of the system, hence potentially increasing the accuracy of the FDC, and thus reducing the complexity of the problem. The consolidation of such expectations would require further analysis, and goes beyond the scopes of this work.

\section{Conclusions}
\label{sec_conclusions}

We have investigated the ability of a GA to reconstruct the topology of a quantum network using indirect probe-based measurements.
While the proposed methodology is entirely general, we have benchmarked its performance by demonstrating that it can effectively reconstruct networks of up to 10 nodes, with performance on larger system sizes limited by the computational requirements of simulating a sufficiently large population.

To enable reconstruction of larger networks, we considered varying the number of measurement configurations used to probe the network. Increasing this number reduces the necessary population size for network reconstruction, hinting at the need for balancing computational and experimental limitations that facilitate the optimal performance of the algorithm depending on the network size.

Our study paves the way for future investigations along multiple directions. An important feature highlighted in our work is the FDC of the fitness function. The latter embodies a simple way to compare the indirect measurements performed upon two networks. Alternative fitness functions, potentially employing machine learning techniques, may facilitate a higher FDC and thus improved performance of the algorithm for larger network sizes. Additionally, there is the possibility to modify the model to relax the constraint of binary connections between nodes and consider reconstruction of networks with different interaction strengths between nodes. Finally, the scenario in which noise is present in the network or imperfect measurements are performed could be considered to investigate  robustness. These remain open questions for future work.

\acknowledgments
C.J.C. and M.M. are grateful to the Quantum Theory group at University of Palermo (https://quantum.unipa.it/) for hospitality during various stages of development of this project.
We acknowledge support from the European Union’s Horizon Europe EIC-Pathfinder
project QuCoM (101046973), the Department for the Economy of Northern Ireland under the US-Ireland R\&D Partnership Programme, the ``Italian National Quantum Science and Technology Institute (NQSTI)" (PE0000023) - SPOKE 2 through project ASpEQCt.

\section*{Conflict of Interest}

The authors have no conflicts to disclose.

\section*{Data Availability}

The data that support the findings of this study are available from the corresponding author
upon reasonable request.

\bibliography{refs.bib}
\bibliographystyle{unsrt}

\end{document}